\documentclass[lettersize,journal]{IEEEtran}
\usepackage{amsmath,amsfonts}
\usepackage{algorithmic}
\usepackage{algorithm}
\usepackage{array}
\usepackage[caption=false,font=normalsize,labelfont=sf,textfont=sf]{subfig}
\usepackage{textcomp}
\usepackage{stfloats}
\usepackage{url}
\usepackage{verbatim}
\usepackage{graphicx}
\usepackage{cite}
\usepackage{balance}
\begin{document}

\title{SDA-DDA: Semi-supervised Domain Adaptation with Dynamic Distribution Alignment Network For Emotion Recognition Using EEG Signals}

\author{Jiahao Tang\textsuperscript{\textdagger}, Chun-Wang Su\textsuperscript{\textdagger}, Youjun Li, Haoyu Wang, Hadia Naeem, Peng Fang, Yangxuan Zheng, Xiangting Fan, Nan Yao, Jue Wang, Xueping Li, Zi-Gang Huang$^{*}$
\thanks{\textsuperscript{\textdagger} These authors contributed equally to this work.}
\thanks{* email: huangzg@xjtu.edu.cn.}
\thanks{This work was supported by the Natural Science Foundation of China (No.11975178), Natural Science Basic Research Program of Shaanxi (No. 2023-JC-YB-07) and Shaanxi Fundamental Science Research Project for Mathematics and Physics (Grant No.22JSQ037)}
\thanks{Jiahao Tang, Chun-Wang Su, Youjun Li, Haoyu Wang, Hadia Naeem, Jue Wang, and Zi-Gang Huang are affiliated with the Key Laboratory of Biomedical Information Engineering, Ministry of Education, Institute of Health and Rehabilitation Science, School of Life Science and Technology, Xi'an Jiaotong University. They are also associated with the Key Laboratory of Neuro-Informatics \& Rehabilitation Engineering, Ministry of Civil Affairs, Xi'an, Shaanxi, China, as well as the Research Center for Brain-Inspired Intelligence, Xi'an Jiaotong University.}
\thanks{Peng Fang is with Department of Military Medical Psychology, AF Medical University, Xian 710032, China.}
\thanks{Nan Yao is with the School of Science, Xi 'an University of Technology, Xi’an, Shaanxi 710048, China. Xueping Li is with the School of Automation and Information Engineering, Xi’an University of Technology, Xi’an, Shaanxi 710048, China.}}
\markboth{Journal of \LaTeX\ Class Files,~Vol.~14, No.~8, August~2021}%
{Shell \MakeLowercase{\textit{et al.}}: A Sample Article Using IEEEtran.cls for IEEE Journals}


\maketitle

\begin{abstract}
In this paper, we focus on the challenge of individual variability in affective brain-computer interfaces (aBCI), which employs electroencephalogram (EEG) signals to monitor and recognize human emotional states, thereby facilitating the advancement of emotion-aware technologies. The variability in EEG data across individuals poses a significant barrier to the development of effective and widely applicable aBCI models. To tackle this issue, we propose a novel transfer learning framework called Semi-supervised Domain Adaptation with Dynamic Distribution Alignment (SDA-DDA). This approach aligns the marginal and conditional probability distribution of source and target domains using maximum mean discrepancy (MMD) and conditional maximum mean discrepancy (CMMD). We introduce a dynamic distribution alignment mechanism to adjust differences throughout training and enhance adaptation. Additionally, a pseudo-label confidence filtering module is integrated into the semi-supervised process to refine pseudo-label generation and improve the estimation of conditional distributions. Extensive experiments on EEG benchmark databases (SEED, SEED-IV and DEAP) validate the robustness and effectiveness of SDA-DDA. The results demonstrate its superiority over existing methods in emotion recognition across various scenarios, including cross-subject and cross-session conditions. This advancement enhances the generalization and accuracy of emotion recognition, potentially fostering the development of personalized aBCI applications.  The source code is accessible at https://github.com/XuanSuTrum/SDA-DDA.
\end{abstract}

\begin{IEEEkeywords}
	Affective brain-computer interfaces, Emotion recognition, Electroencephalogram, Individual variability, Transfer learning, Semi-supervised domain adaptation, Dynamic distribution alignment
\end{IEEEkeywords}

\section{Introduction}
Emotions are crucial to the human experience, exerting profound influence on cognitive processes and decision-making. They deeply affect human thoughts shaping, behaviors, and interpersonal relationships. The accurate monitoring, recognition, and regulation of emotions have profound influence in improving the lives of individuals, and also optimize the performance of specific task operators.
Various characteristics have been adopted in recognizing human emotional states, such as facial expression, gestures, tone of voice, and physiological signals\cite{hernandez2016facial,singh2021multimodal}. Among these, physiological signals have gained widespread adoption for their objectivity of assessment. Discriminating emotional states with physiological signals can be achieved through various techniques, including ECG, EMG, EOG, respiration rate, and EEG\cite{duncombe1959infrared}. EEG, known for recording the brain's electrical activity, is commonly employed in emotion monitoring via affective brain-computer interfaces (aBCI), attracting considerable academic interest.\cite{wu2023affective}.

Presently, the inherent variations in individuals’ physiological structures and brain activities present significant challenges in the development of affective recognition systems based on EEG signals. These variations lead to psychological states, emotional responses, and regulation strategies when individuals face emotional stimuli. Developing a model capable of adapting to individual differences with limited labeled data remains a major obstacle. A promising approach to tackle this challenge involves utilizing transfer learning techniques.\cite{wu2020transfer}. Transfer learning leverages knowledge obtained from one dataset or domain (commonly referred to as the source domain ($D_s$)) to improve the performance of learning tasks in a different dataset or domain (referred to as the target domain ($D_t$)). Specifically, in EEG-based emotion detection, the source domain typically consists of labeled EEG signals collected from multiple participants, whereas the target domain includes unlabeled EEG data\cite{pan2009survey}. By incorporating transfer learning methods, the disparity in individual differences can be mitigated, facilitating improved emotional brain-computer interfaces\cite{wan2021review}. This methodology enables the model to utilize prior knowledge extracted from the source domain, which comprises data from diverse subjects, to enhance its ability to interpret emotional states in the target domain. This enhancement proves particularly valuable when the target domain contains only unlabeled EEG data\cite{li2021can}.

Domain adaptation, a critical area within transfer learning, provides an effective approach to bridging the disparities in data distributions between $D_s$ and $D_t$ datasets\cite{pan2009survey}. In domain adaptation, it is assumed that the feature space, conditional probability distribution(CPD) and label space of both domains remain consistent, while variations exist in their margin probability distribution(MPD). 

Based on these assumptions and methodologies, researchers have successfully addressed the issue of individual differences between the source and target domains in emotional brain-computer interfaces\cite{kouw2019review,guan2021domain,pan2023st}. Within the domain of EEG emotion recognition, multiple strategies have been developed to tackle challenges associated with domain adaptation and feature extraction. One noteworthy approach is multi-source marginal distribution adaptation (MS-MDA), which assumes that EEG data across various sources share low-level characteristics. It uses separate branches for each source domain to enable one-to-one domain alignment and to extract features specific to each domain\cite{chen2021ms}. Another prominent method, adversarial discriminative temporal convolutional networks (AD-TCN), focuses on domain adaptation by maintaining consistent graph-based feature representations across domains while simultaneously capturing domain-specific variations\cite{he2022adversarial}. Collectively, these approaches aim to address the challenges of domain adaptation and feature extraction, contributing to the advancement of EEG affective recognition systems.

While these methods show promise, the majority of research on EEG-based emotion recognition has focused primarily on aligning the MPD across domains, often neglecting the alignment of the CPD. Aligning the MPD alone fails to account for the intrinsic variations that exist within each emotion category across individuals. This limitation restricts the model’s ability to capture discriminative, domain-invariant features that are crucial for accurate emotion recognition. As a result, the alignment of CPDs within emotional categories continues to be a critical but largely overlooked challenge in the field.

This paper introduces a novel transfer learning framework, SDA-DDA, specifically designed to enhance cross-subject emotion recognition using EEG dataset. SDA-DDA consists of four distinct modules: MPD alignment via maximum mean discrepancy, CPD alignment via conditional maximum mean discrepancy, joint distribution dynamic adjustment, and a semi-supervised pseudo-label optimization algorithm. The key contributions and innovations presented in this study are summarized below:

(1)Our framework advances EEG-based emotion recognition by dynamically aligning both MPD and CPD between source and target domains. Contrary to traditional methods which use static alignment, SDA-DDA’s dual distribution alignment approach continuously adapts to changes in both types of distributions throughout training. MMD handles marginal distribution alignment using kernelized moment matching, while CMMD is employed to minimize differences in conditional distributions for each emotion category. This dual approach enhances the model’s adaptability to individual variations among subjects, providing a more robust solution for cross-subject emotion recognition.

(2)To address the challenges of semi-supervised learning with EEG data, we introduce a dynamic confidence threshold adjustment mechanism. Pseudo-labels are generated from unlabeled data during training. To maintain the quality of pseudo labels, only high-confidence samples are selected to enhance the learning process. By dynamically adjusting this confidence threshold, our model selectively filters pseudo-labels to improve the quality of semi-supervised learning, making it more robust to noise in the unlabeled data and enhancing the overall performance of the affective brain-machine interface.

In summary, the proposed SDA-DDA framework solves the limitations of existing methods by introducing dual distribution alignment and a dynamic confidence-based filtering mechanism. These innovations enable SDA-DDA to more effectively handle the unique challenges of cross-subject emotion recognition with EEG data, leading to improved classification accuracy and stability in affective brain-machine interfaces. Notations and descriptions used in this paper is shown in Table~\ref{tab1}

\begin{table}[h!]
	\begin{center}
		\caption{Notations and descriptions used in this paper.}\label{tab1}
		\setlength{\tabcolsep}{1mm}
		\begin{tabular}[t]{l|c}
			\hline
			Notation & Description \\
			\hline
			$D_s=\left\{ x_{s}^{l},y_{s}^{l} \right\} _{i=1}^{N_{}}$ &Source domain\\
			$D_t=\left\{ x_{t}^{u} \right\} _{i=1}^{Nu}$&Target domain \\
			$f()$& Feature extractor \\
			$\hat{y}_{s}^{l}$ & Source domain predict labels \\
			$\hat{y}_{t}^{u}$ & Target generate pseudo labels\\
			$K$ & Gaussian Kernel function \\
			$RKHS$ & Reproducing Kernel Hilbert Space\\
			$SGD$  & Stochastic Gradient Descent\\
			$ReLU$ & Rectified Linear unit activation function\\
			\hline
		\end{tabular}
	\end{center}
\end{table}

\section{Background}\label{sec:background}
\subsection{EEG Emotion Recognition with Deep Learning}
Many studies have used classical machine learning methods to realize EEG-based emotional state recognition. These methods mainly involve two typical steps: feature extraction and classifier classification. However, machine learning approaches for emotional EEG signal recognition have several drawbacks, such as their high-dependence on manual feature engineering, the requirement for domain expertise to extract relevant features from EEG signals, and limited applicability in large-scale data settings due to a lack of automatic feature learning\cite{ bazgir2018emotion}.

Advancements in research have propelled deep learning techniques into widespread use, particularly in the field of affective detection. These methods exhibit remarkable performance across diverse tasks, particularly when utilized for analyzing EEG signals\cite{craik2019deep}. In contrast to traditional machine learning methods, deep learning offers significant advantages, such as enhanced learning capabilities and adaptability for handling extensive datasets. One proposed approach focuses on an emotion detection method leveraging multi-channel EEG data. This technique creates a three-dimensional representation by combining spatial and spectral features. By integrating a feature fusion module with convolutional neural networks, it achieves notable accuracy in emotion identification, with classification rates of 89.67\% for arousal and 90.93\% for valence\cite{yao2022feature}. Another innovative model addresses the limitations of low accuracy in emotion recognition systems for brain-computer interfaces. This approach incorporates the intricate characteristics of EEG signals, accounting for the brain's spatial organization and the temporal aspects of emotional states. EEG feature tensors, categorized by brain regions, are subsequently processed through a hybrid architecture combining convolutional neural networks with bidirectional long short-term memory networks. This structure effectively enhances the system's recognition accuracy. Simulation results further validate the model’s performance, achieving an average accuracy of 94\% on the DEAP dataset and 94.82\% on the SEED dataset\cite{cuinovel}.

The methods mentioned above have demonstrated impressive accuracy in intra-individual affective detection. However, EEG-based affective detection models which achieved high accuracy with subjects may not generalize well on cross-subjects tasks.

\subsection{EEG Emotion Recognition with semi-supervised domain adaptation}
In real-world scenarios, obtaining fully labeled datasets poses a significant challenge, often leading to a large proportion of unlabeled data. Semi-supervised learning addresses this issue by acting as an intermediary between supervised and unsupervised approaches. By leveraging a small subset of labeled data in conjunction with a substantial amount of unlabeled data during training \cite{yang2022survey}, semi-supervised learning emerges as a promising solution for scenarios with limited labeled information. 

In the context of EEG-based affective detection, numerous models employ semi-supervised domain adaptation methods. Among these, domain adversarial neural networks and deep domain confusion (DDC) techniques have shown considerable success in earlier studies. The DDC method utilizes a deep convolutional neural network (DCNN) as its foundational model and incorporates a domain confusion layer to minimize discrepancies in feature distributions between source and target datasets. By simultaneously optimizing domain confusion and classification losses, the DDC approach enables the extraction of robust, domain-invariant features that remain unaffected by domain variations\cite{tzeng2014deep}. This methodology equips the model to generalize effectively to new domains, making it suitable for both supervised and unsupervised adaptation tasks.

Researchers have also explored other approaches to realizing domain invariance of encoded representations. For example, a multi-source learning architecture utilizing the maximum mean discrepancy loss has been proposed, which aligns domains through dataset-specific private encoders\cite{bethge2022domain}. Additionally, an adversarial domain adaptation technique utilizing a multi-branch capsule network (DA-CapsNet) is employed to further reduce discrepancies between the data distributions of source and target domains. This enhancement significantly improves cross-subject emotion recognition performance\cite{liu2024capsnet}. To further advance domain adaptation in emotion recognition, researchers have explored integrated frameworks that combine MPD and CPD. For instance, researcher proposed a unified framework, achieving joint probability distribution adaptation training without the need for $D_t$ label\cite{jimenez2023cross}. 

In contrast to existing methods, our proposed SDA-DDA framework offers several distinctive innovations that optimize cross-subject emotion recognition in EEG-based applications. First, SDA-DDA removes the need for complex adversarial training strategies and complicated feature extraction steps, making the framework more computationally efficient and easier to implement. Second, SDA-DDA employs a dynamic distribution adaptation mechanism that continuously adjusts according to discrepancies between MPD and CPD during model training. This adaptability enables the model to effectively capture the variable relationship between these distributions, improving its ability to generalize across subjects. Finally, our approach incorporates a pseudo-label filtering mechanism based on a dynamic confidence threshold. This mechanism systematically filters pseudo-labels to retain only the most reliable samples, thereby enhancing the accuracy of conditional distribution alignment and reinforcing the discriminative capacity of learned features. Collectively, these features make SDA-DDA a robust, scalable approach for emotion recognition through domain adaptation, offering significant advancements over existing methods.
\begin{figure*} 
	\centering
	\includegraphics[width=0.9\linewidth]{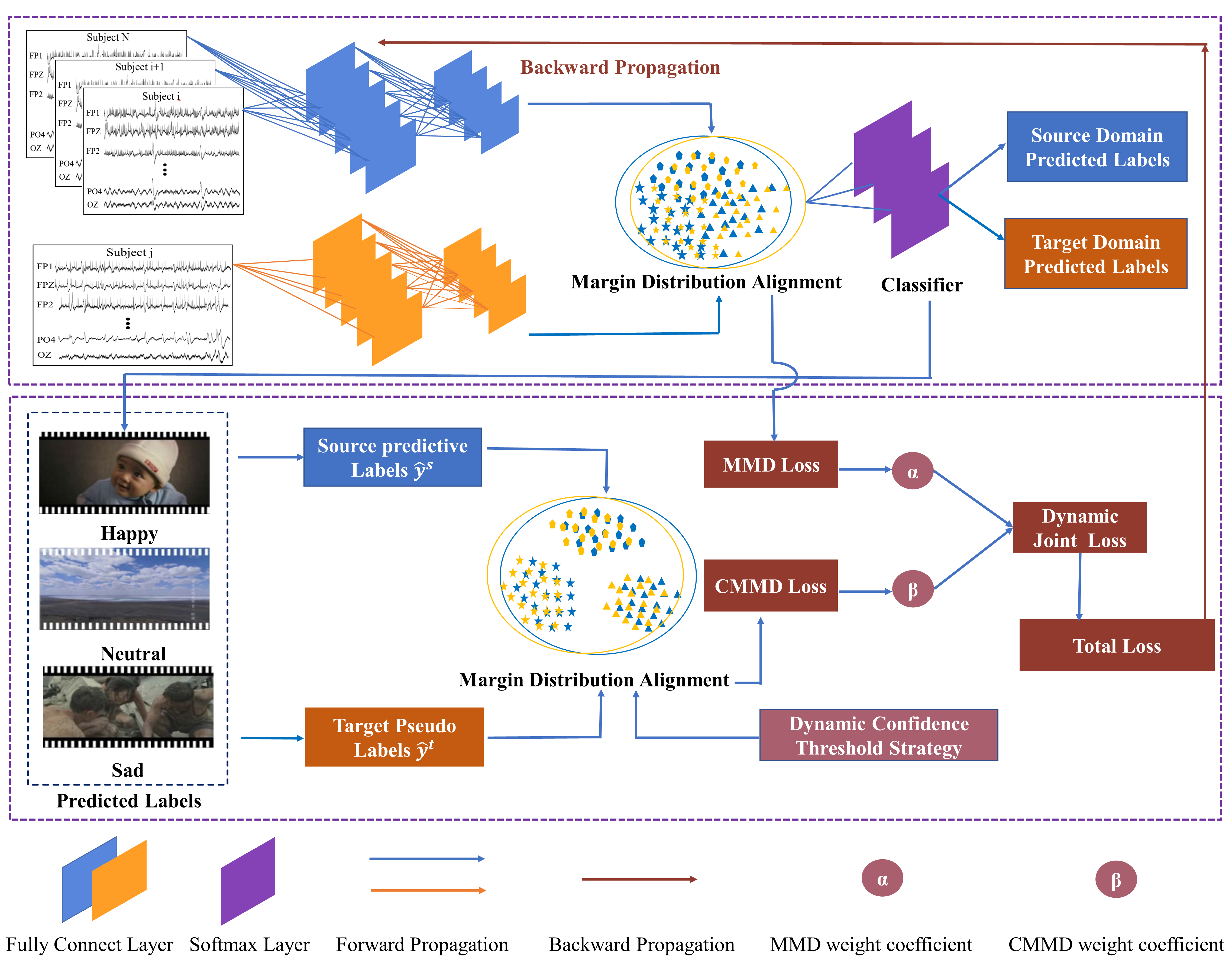}
	\caption{The flowchart of the proposed SDA-DDA framework.} \label{fg:method}
\end{figure*} 
\section{The Proposed Approach}\label{sec:approach}
This paper presents a novel learning framework for cross-subject affective detection based on EEG data. The framework consists of four key modules: The MPD alignment based on maximum mean discrepancy, the CPD alignment based on conditional maximum mean discrepancy, the joint distribution dynamic adjustment, and a semi-supervised pseudo-label optimization algorithm.

As depicted in Fig. 1, the proposed framework begins by extracting features from the $D_s$ and $D_t$, aligning their MPDs using the MMD algorithm. A classifier is then employed to generate label predictions for both domains. To enhance performance, the framework minimizes cross-entropy loss between the predicted and actual labels in the source domain. Additionally, it incorporates a loss term addressing the CPD of the predicted labels in the $D_t$, ensuring effective domain adaptation. During training, the model dynamically adjusts the weights for both MPD and CPD distributions, considering their individual contributions to the overall adaptation process. Furthermore, a screening mechanism is employed to improve the quality of pseudo-labels assigned to the $D_t$. The entire framework is trained using backward propagation, minimizing three aforementioned loss components. This training process enables adaptability adjustments and optimizes the classification performance of the model, facilitating accurate cross-subject emotion recognition based on EEG data. 

\subsection{Feature extraction}
In the domain of EEG signal analysis, a meticulous process of feature extraction was conducted. Firstly, a specific time segment of interest was carefully selected from the EEG signal to capture relevant temporal information effectively. Subsequently, the short-time Fourier transform (STFT) was applied to decompose the EEG signal into five distinct frequency bands: $\delta$, $\theta$, $\alpha$, $\beta$, and $\gamma$. Within each frequency band, the probability density function of the signal was computed to reveal the distribution of signal amplitudes.

The core of this feature extraction process lies in the calculation of the differential entropy, a metric that quantifies the level of uncertainty and complexity inherent in the EEG signal. The differential entropy was computed using the following formula:
\begin{equation}\label{entropy}
	H\left( x \right) =\mathrm{ }-\int{p(x)\log p(x) dx}
\end{equation}

In Eq.(\ref{entropy}), $H(X)$ represents the differential entropy, $p(x)$ denotes the probability density function that characterizes the amplitude distribution of the signal within the specific frequency band, and $x$ represents the amplitude of the signal.

This feature extraction procedure was systematically applied to each of the frequency bands, calculating differential entropy values for each individual band. Consequently, each EEG sample is associated with a comprehensive 310-dimensional differential entropy feature vector, which incorporates information from the five frequency bands across the 62 EEG channels. This feature representation provides a rich and informative basis for in-depth analysis and interpretation of EEG data.

\subsection{Alignment of Marginal Probability Distribution}
In this model, aligning the MPDs of the $D_s$ and $D_t$ is crucial. To measure the divergence between these distributions, the Maximum Mean Discrepancy is utilized as a key metric. The MMD-based distance is formally defined as follows:

\begin{equation}\label{MMD1}
	MMD\left( D_s,D_t \right) =||\frac{1}{n}\sum_{i=1}^n{\varPhi \left( x_{s}^{_l} \right)}-\frac{1}{m}\sum_{j=1}^m{\varPhi \left( x_{t}^{u} \right)}||_{H}^{2}
\end{equation}

Here, $MMD\left( D_s,D_t \right)$ represents the MMD-based distance between the $D_s$ and the $D_t$. The measurement of this distance is carried out via the $\varPhi$ function, which skillfully maps the data from both domains into a RKHS denoted as $H$. 
According to the theorem $||\mathbf{A}||^2 = \operatorname{tr}(\mathbf{A} \mathbf{A}^{\mathrm{T}})$ and $\operatorname{tr}(\mathbf{A} \mathbf{B}) = \operatorname{tr}(\mathbf{B} \mathbf{A})$, equation (\ref{MMD1}) can be rewritten as follows:

\begin{equation}\label{MMD2}
	\begin{split}
		&||\frac{1}{n}\sum_{i=1}^n{\varPhi \left( x_{s}^{_l} \right)}-\frac{1}{m}\sum_{j=1}^m{\varPhi \left( x_{t}^{u} \right)}||_{H}^{2}\\
		&= 
		tr\left( \frac{1}{n^2}\varPhi \left( x_{s}^{l} \right) \varPhi \left( x_{s}^{l} \right) ^T 
		+ \frac{1}{n^2}\varPhi \left( x_{t}^{u} \right) \varPhi \left( x_{t}^{u} \right) ^T \right. \\
		&\quad \left. - \frac{1}{n^2}\varPhi \left( x_{s}^{l} \right) \varPhi \left( x_{t}^{u} \right) ^T 
		- \frac{1}{n^2}\varPhi \left( x_{t}^{u} \right) \varPhi \left( x_{s}^{l} \right) ^T \right) \\
		&=\operatorname{tr} \left( \begin{bmatrix}
			\varPhi \left( x_{s}^{_l} \right) & \varPhi \left( x_{t}^{u} \right)
		\end{bmatrix} \begin{bmatrix}
			\frac{1}{n^2} \mathbf{1}\mathbf{1}^{\mathrm{T}} & \frac{-1}{n m} \mathbf{1}\mathbf{1}^{\mathrm{T}} \\ 
			\frac{-1}{n m} \mathbf{1}\mathbf{1}^{\mathrm{T}} & \frac{1}{m^2} \mathbf{1}\mathbf{1}^{\mathrm{T}}
		\end{bmatrix} \begin{bmatrix}
			\varPhi \left( x_{s}^{_l} \right)^\mathrm{T} \\ \varPhi \left( x_{t}^{u} \right)^\mathrm{T}
		\end{bmatrix} \right)\\
		&=\operatorname{tr} \left( \begin{bmatrix}
			\varPhi \left( x_{s}^{_l} \right) \\ \varPhi \left( x_{t}^{u} \right)^\mathrm{T}
		\end{bmatrix}  \begin{bmatrix}
			\varPhi \left( x_{s}^{_l} \right) & \varPhi \left( x_{t}^{u} \right)
		\end{bmatrix} \begin{bmatrix}
			\frac{1}{n^2_s} \mathbf{1}\mathbf{1}^{\mathrm{T}} & \frac{-1}{n_s n_t} \mathbf{1}\mathbf{1}^{\mathrm{T}} \\ 
			\frac{-1}{n_s n_t} \mathbf{1}\mathbf{1}^{\mathrm{T}} & \frac{1}{n^2_t} \mathbf{1}\mathbf{1}^{\mathrm{T}}
		\end{bmatrix} \right)\\
		&=\operatorname{tr} \left( \begin{bmatrix}
			<\varPhi \left( x_{s}^{_l} \right),\varPhi \left( x_{s}^{_l} \right)> & <\varPhi \left( x_{s}^{_l} \right),\varPhi \left( x_{t}^{u} \right)>\\ 
			<\varPhi \left( x_{t}^{u} \right),\varPhi \left( x_{s}^{_l} \right)> & <\varPhi \left( x_{t}^{u} \right),\varPhi \left( x_{t}^{u} \right)>
		\end{bmatrix} \mathbf{M} \right)\\
		&=\mathrm{tr}\left( \left[ \begin{matrix}
			K\left( x_{s}^{l},x_{s}^{l} \right)&		K\left( x_{s}^{l},x_{t}^{u} \right)\\
			K\left( x_{t}^{u},x_{s}^{l} \right)&		K\left( x_{t}^{u},x_{t}^{u} \right)\\
		\end{matrix} \right] \mathbf{M} \right) 
	\end{split}
\end{equation}
Where $\left(M\right)_{i j}=\left\{\begin{array}{ll}{\frac{1}{n^2},} & {\mathbf{x}_{i}, \mathbf{x}_{j} \in \mathcal{D}_{s}} \\ {\frac{1}{m^2},} & {\mathbf{x}_{i}, \mathbf{x}_{j} \in \mathcal{D}_{t}} \\ {\frac{-1}{n m},} & {\text { otherwise }}\end{array}\right.$

The MMD is realized through Eq. (\ref{MMD2}) as presented below:

\begin{equation}\label{MMD2}
	MMD\left( D_s,D_t \right) =\frac{1}{n^2}K_{s,s}+\frac{1}{m^2}K_{t,t}-\frac{2}{nm}K_{s,t}
\end{equation}

We employ the Gaussian kernel function, denoted as $k\left( u,v \right) =e^{\frac{-\left\| u-v \right\| ^2}{\sigma}}$, as our kernel function. The Gaussian kernel function and the MMD metric facilitate the efficient evaluation of distributional dissimilarity between the $D_s$ and $D_t$.

\subsection{Alignment of Conditional Probability Distribution}
In this section, we address a crucial aspect often neglected in the majority of transfer learning studies with respect to EEG signals. These studies predominantly concentrate on aligning MPDs but tend to overlook the pivotal role played by aligning CPD in overall data harmonization. To address this gap, we introduce a novel approach termed conditional maximum mean discrepancy. CMMD is specifically designed to facilitate alignment of CPD across distinct emotional categories within EEG data.

Effective alignment of CPD plays an important role in achieving robust distributional adaptation. However, when dealing with $Q_t\left(y_{t}^{u}|x_{t}^{u} \right) $ and $P_s\left( y_{s}^{l}|x_{s}^{l} \right)$, we encounter two significant challenges. Firstly, obtaining the posterior probability distribution presents a formidable task. Moreover, the lack of labels in the $D_t$ adds another layer of complexity to the problem. In order to address these issues, this paper introduces two key assumptions to guide our approach:

\begin{enumerate}
	\item To mitigate the lack of labeled data in the $D_t$, we introduce an approach based on pseudo-labeling. Here, the deep neural network's outputs, denoted as $\hat{y}_t^u = f(x_t^u)$, act as pseudo-labels for $D_t$ samples. The reliability of these pseudo-labels increases incrementally during the iterative process.
	\item Given the complexities in estimating posterior probabilities $Q_t\left(y_{t}^{u}|x_{t}^{u} \right) $ and $P_s\left( y_{s}^{l}|x_{s}^{l} \right)$, we shift our focus towards exploring the sufficient statistics of class-conditional distributions, namely, $P(x_{s}^{l}|y_{s}^{l} = c)$ and $Q(x_{t}^{u}|y_{t}^{u} = c)$ for each class $c \in \{1,\dots,C\}$. This allows us to effectively match class-conditional distributions across $D_s$ and $D_t$.
\end{enumerate}
\begin{figure} 
	\centering
	\includegraphics[width=0.9\linewidth]{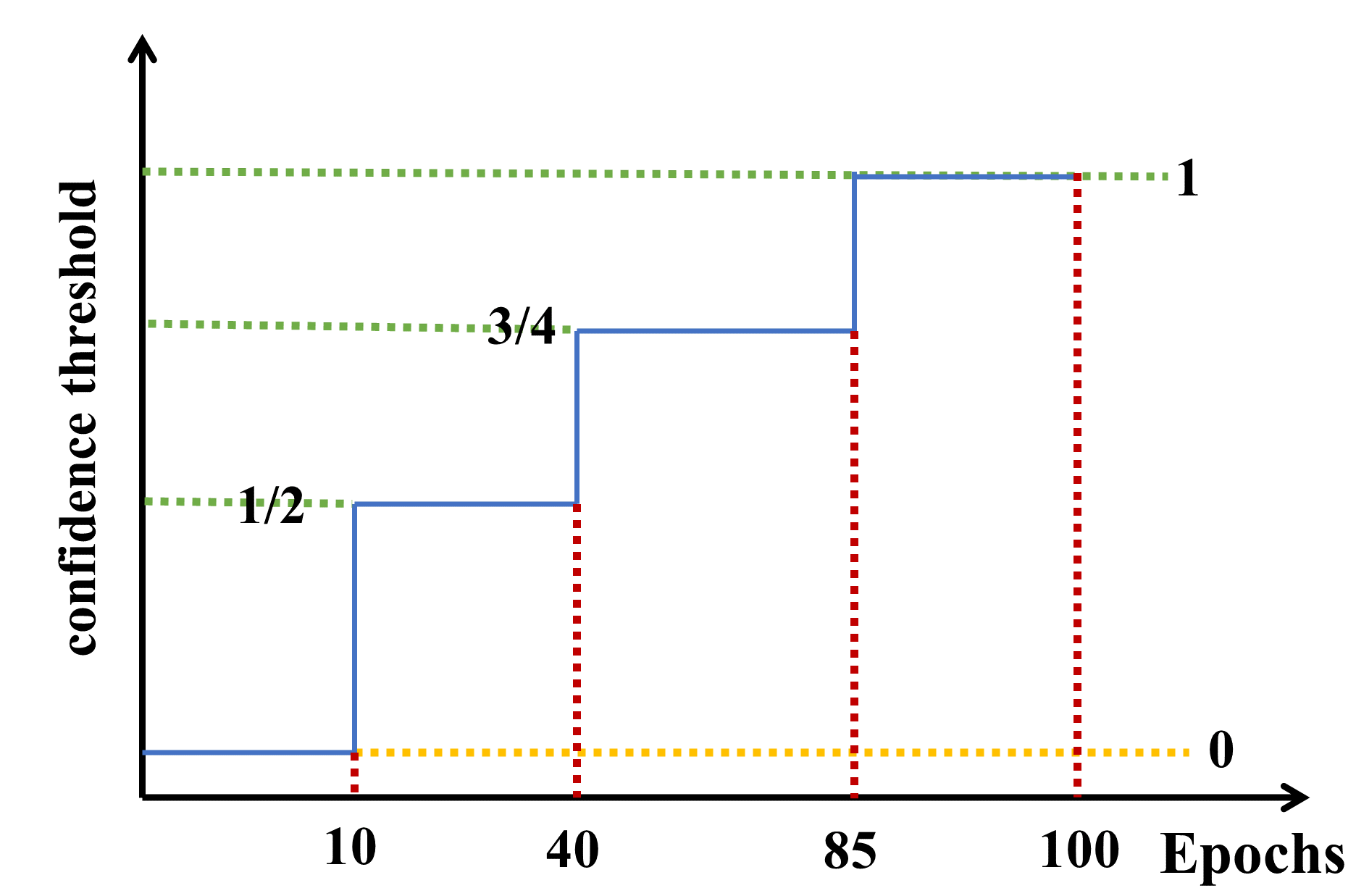}
	\caption{The pseudo labels confidence threshold mechanism.} \label{fg:Pse-con}
\end{figure}

\begin{equation}\label{CMMD}
	\begin{split}
		&CMMD\left( D_s,D_t \right) \\
		&= 
		\frac{1}{C}\sum_{c=1}^C{\left\| \frac{1}{n_{s}^{c}}\sum_{i=1}^{n_{s}^{c}}{\varPhi \left\{ \left( x_{s}^{l} \right) ^c \right\} -\frac{1}{n_{t}^{c}}\sum_{j=1}^{n_{t}^{c}}{\varPsi \left\{ \left( x_{t}^{u} \right) ^c \right\} }} \right\|}_{H}^{2}\\	
		&=
		\frac{1}{C}\sum_{c=1}^C{\frac{1}{\left( n_{s}^{c} \right) ^2}}\sum_{i=1}^{n_{s}^{c}}{\sum_{j=1}^{n_{s}^{c}}{K_{s,s}}} \\
		&+
		\frac{1}{C}\sum_{c=1}^C{\frac{1}{\left( n_{t}^{c} \right) ^2}}\sum_{i=1}^{n_{t}^{c}}{\sum_{j=1}^{n_{t}^{c}}{K_{t,t}}} \\
		& -\frac{2}{C}\sum_{c=1}^C{\frac{1}{n_{s}^{c}n_{t}^{c}}\sum_{i=1}^{n_{s}^{c}}{\sum_{j=1}^{n_{t}^{c}}{K_{s,t}}}}
	\end{split}
\end{equation}
Assuming that $\varPhi$ and $\varPsi$ represent nonlinear mappings of $D_s$ and $D_t$, respectively, we utilize conditional kernel mean embedding to project the CPD into a series of points in the RKHS. The conditional kernel mean embedding is defined as Eq. (\ref{CMMD}). In this context, $c$ signifies the $c$-th category among the $C$ labels, where $n_{s}^{c}$ and $n_{t}^{c}$ indicate the number of samples belonging to category $c$ in the source and target datasets, respectively. Additionally, $\left( x_{s}^{l} \right) ^c$ corresponds to the $i$-th labeled sample in the $D_s$ datasets for category $c$, and $\left( x_{t}^{u} \right) ^c$ corresponds to the $i$-th pseudo-labeled sample in the $D_t$ datasets for the same category.

As depicted in Fig.~\ref{fg:method}, the $D_t$ undergoes classification to obtain pseudo-labels, and the results of these pseudo-labels are input into the CMMD computation formula (Eq. (\ref{CMMD})). Therefore, the quality of pseudo-labels significantly affects the efficiency of aligning CPD.

To enhance the performance of domain adaptation learning and allow the model to selectively retain $D_t$ samples with high classification confidence across different training stages, this paper proposes a dynamic confidence threshold strategy. High-quality pseudo-labels $\hat{D}_t$ can be selected using Eq. (\ref{pse-s}).

\begin{equation}\label{pse-s}
	\hat{D}_t=\left\{ \mathcal{\varGamma} \left( p_{i}^{u},\tau \right) \hat{y}_{t}^{u} \right\} _{i=1}^{N_u}
\end{equation}
where $\tau \in \left[ 0,1 \right] $, $ \mathcal{\varGamma} \left( \cdot \right) $ is confidence threshold mechanism. The $ \mathcal{\varGamma} \left( \cdot \right)$ function is illustrated in Fig.~\ref{fg:Pse-con}. The strategy is contingent upon the training stage, dynamically adjusting the confidence threshold accordingly. In the early stages ($0 \le \text{epoch} < 10$), a low threshold of 0 is set to fully includes all $D_t$ samples. During the training progresses, the threshold gradually increases to 0.5 in the middle stage ($10 \le \text{epoch} < 40$) and further to 0.75 in the late stage ($40 \le \text{epoch} \le 85$), aiming to selectively retain $D_t$ samples with higher classification confidence. Beyond this range, the threshold is set to 1, indicating a requirement for higher confidence in retaining samples.


\subsection{Loss function and Dynamic adjustment factor}
The framework proposed in this study considers three distinct loss functions: the classification loss for the $D_s$, the loss associated with MPD disparities, and the loss for differences in CPD.

The proposed framework primarily focuses on sample differentiation by optimizing the label loss. Specifically, it aims to minimize the cross-entropy loss between the true labels and the predicted labels of $D_s$ samples. The loss function can be defined as follows:

\begin{equation}\label{loss1}
	\mathcal{L} _{Ds}=-\frac{1}{B_L}\sum_{i=1}^{B_L}{\sum_{c=1}^C{y_{s}^{l}\log \left( \left( p_{s}^{l} \right) ^c \right)}}
\end{equation}

Here, $N$ represents the total number of samples, and $C$ stands for the number of categories. In the context of this equation, $y_{s}^{l}$ signifies the value located at the $i$-th position within the ground-truth label (in the form of a one-hot vector), while $ \left( p_{s}^{l} \right) ^c$ represents the predicted probability that the observation sample $i$ belongs to category $c$.

As depicted in Fig.~\ref{fg:method}, following the computation of MMD loss and CMMD loss, the relative significance of the MPD and CPD is dynamically adjusted through the introduction of weighting parameters $\alpha$ and $\beta$. 

The parameter $\alpha$ adjusts the weight of the MMD loss throughout the training process. During the early stages of training, the model ensures comprehensive capture of the MPD differences between the $D_s$ and $D_t$ by setting the MMD weight to $\tau_h$. As the training progresses iteratively, the MMD weight is gradually reduced to balance its contribution with CMMD, allowing the model to adapt to the $D_t$. As training progresses and the model reaches greater stability, decreasing the MMD weight to $\tau_l$ increasingly emphasizes CPD alignment, mitigating the risk of overfitting of the $D_s$ to the $D_t$. The weighting parameter $\beta$ is given by:

\begin{equation}\label{beta}
	\beta \,\,=\varepsilon \left( \rho _0-\mathcal{L} _{Ds} \right) +\frac{1}{2}\varepsilon \left( \mathcal{L} _{Ds}-\rho _0 \right) \varepsilon \left( \rho _1-\mathcal{L} _{Ds} \right)
\end{equation}

The $\varepsilon$ represents the step function. represents the step function. The above formulas dynamically adjusts the weighting factor of the CPD. 
The minimization of the above equations leads to a reduction in the CPD between the $D_s$ and $D_t$, thereby contributing to the alignment of the CPD..
\begin{equation}\label{loss3}
	\begin{split}
		\mathcal{L} =\mathcal{L} _{Ds}+\alpha \mathcal{L} _{mmd}+\beta \mathcal{L} _{cmmd}
	\end{split}
\end{equation}

The overall algorithm of this paper is summarized in Algorithm~\ref{alg:AOS}.

\begin{algorithm}[!h]
	\caption{: Algorithm of the SDA-DDA model}\label{alg:AOS}
	\renewcommand{\algorithmicrequire}{\textbf{Input:}}
	\renewcommand{\algorithmicensure}{\textbf{Output:}}
	
	\begin{algorithmic}[1]
		\REQUIRE $D_s=\left\{ x_{s}^{l},y_{s}^{l} \right\} _{i=1}^{N_{}}$ $D_t=\left\{ x_{t}^{u} \right\} _{i=1}^{Nu}$
		\ENSURE The predicted label of the target domain    
		
		\STATE Random a mini-batch $D_{batch}^{s}\,\, and D_{batch}^{t}\,\,$ from $D_s,D_t$ respectively;
		\STATE Extract the EEG common features and Calculate the MMD loss by Eq.(\ref{MMD1});
		\STATE Generate predict pseudo-labels for each sample $x_{t}^{u}\in D_{batch}^{t}$ by classifier;
		\STATE Pseudo-labels are filtered using a dynamic confidence threshold adjustment mechanism during the training process by Eq.(\ref{pse-s}).
		\STATE Estimate the CMMD loss and cross-entropy loss by Eqs. (\ref{CMMD}) and (\ref{loss1}) respectively;
		\STATE Update the network parameters by gradient descend to minimize Eq.(\ref{loss3})\\
		\RETURN The predicted label of the $D_t$
	\end{algorithmic}
\end{algorithm}
\section{Experiments and Results}\label{sec:experimentresults}
\subsection{Emotion datasets}
The primary datasets utilized in this study are SEED and SEED\_IV, which are publicly accessible through the Brain-like Computing and Machine Intelligence Center at Shanghai Jiao Tong University. In the SEED dataset, experiments were conducted with fifteen Chinese participants, each completing three sessions comprising a total of fifteen trials. Participants were exposed to Chinese film clips designed to evoke a range of emotional states, including positive, neutral, and negative emotions\cite{zheng2015investigating}. In the SEED\_IV dataset, experiments were conducted with 15 subjects across three sessions on separate days, with each session including 24 trials.Participants in each trial watched film clips designed to elicit emotions such as happiness, sadness, neutrality, or fear. EEG data for the SEED and SEED\_IV datasets were recorded with a 62-channel ESI NeuroScan System.\cite{8283814}.

In the DEAP dataset, 32 participants (equally split between male and female) viewed 40 one-minute music videos while their EEG signals were recorded from 32 electrodes. Post-viewing, they rated arousal, valence, dominance, and liking using self-assessment manikins on a continuous scale from 1 to 9\cite{koelstra2011deap}

\subsection{Implementation details}
In the model presented in this paper, the feature extractor is composed of a pair of fully connected layers. The initial layer, denoted as "fc1," processes input data spanning 310 dimensions and reduces it to 64 dimensions using a ReLU activation function. Subsequently, the data flows through the second layer, labeled "fc2," further reducing the dimensionality to 64, accompanied by an additional ReLU activation. To improve performance and prevent overfitting, dropout layers with a 0.25 dropout rate are applied after each fully connected layer. For the parameters $\alpha$, $\tau_h$ and $\tau_l$are 1, 0.01, respectively. For the Eq. (\ref{beta}) , $\rho _0$ and  $\rho _1$ are 0.1 and 0.15, respectively.

For our experimental setup, we utilize all labeled source samples to represent the $D_s$, while the unlabeled target samples constitute the $D_t$. We evaluate and compare the average classification accuracy as a performance measure. Our training configuration encompasses several essential parameters and procedural steps. The batch size for training is configured to 32, and the training process spans 10 epochs. We employ two distinct learning rates, specifically 0.001 and 0.01. The SGD momentum parameter is set to 0.9. Moreover, the option to enable or disable CUDA training is provided to the user. Additionally, we initialize a random seed value of 3 to ensure result reproducibility, and the L2 weight decay is precisely set at $5 \times 10^{-4}$. During training, we employ the SGD optimization method, with learning rates subject to dynamic adjustments at the beginning of each epoch. These adaptive changes in learning rates follow a specific mathematical formula. This iterative training process continues for the designated number of epochs, allowing the model to learn and adapt over time.

\begin{table}
	\centering
	\caption{The Results of representation methods on SEED datasets using Cross-Subject Single-Session Leave-One-Subject-Out Cross-Validation}
	\setlength{\tabcolsep}{1mm}
	\label{tab2}
	\begin{tabular}{c|c|c|c}
		\hline
		Method & Pcc(\%) & Method & Pcc(\%) \\
		\hline
		KPCA~\cite{li2018cross} & 61.28$\pm$14.62 & TPT~\cite{li2018cross} & 75.17$\pm$12.83  \\\hline              
		ADA~\cite{li2019domain} & 84.47$\pm$10.65 & DANN~\cite{li2019domain} & 81.65$\pm$09.92 \\\hline
		MFA-LR~\cite{jimenez2023learning} & 85.27$\pm$10.84 & DA-CapsNet~\cite{liu2024capsnet} &84.63$\pm$09.09 \\\hline
		MS-MDA~\cite{chen2021ms} & 89.63$\pm$06.97 & SCSTM-DS~\cite{chen2023similarity} & 82.29 $\pm$03.60\\\hline
		GCPL~\cite{li2024generalized} & 80.74$\pm$06.05 & MS-ADA~\cite{she2023multisource} & 86.16 $\pm$07.87\\\hline
		MSFR-GCN~\cite{pan2023msfr} & 86.78$\pm$05.40 & \textbf{SDA-DDA}  & 87.27$\pm$07.55\\\hline
	\end{tabular} 
\end{table}

\begin{table}
	\centering
	\caption{The performance of representation methods on SEED\_IV datasets using Cross-Subject Single-Session Leave-One-Subject-Out Cross-Validation}
	\setlength{\tabcolsep}{1mm}
	\label{tab3}
	\begin{tabular}{c|c|c|c}
		\hline
		Method & Pcc(\%) & Method & Pcc(\%) \\
		\hline
		MS-MDA~\cite{chen2021ms} & 59.34$\pm$05.48& MS-STM~\cite{li2019multisource} & 61.41$\pm$09.72 \\\hline
		MS-ADRT~\cite{jiang2023generalization} & 68.98$\pm$06.80  &  DANN~\cite{li2019domain} & 54.63$\pm$08.03 \\\hline
		GCPL~\cite{li2024generalized} & 62.65$\pm$09.79 & MS-ADA~\cite{she2023multisource} & 59.29 $\pm$13.65\\\hline
		MSFR-GCN~\cite{pan2023msfr} & 73.43$\pm$07.32 & \textbf{SDA-DDA}  & 74.01$\pm$11.34\\\hline
	\end{tabular}
\end{table}

\subsection{Experiment Setting}
In this study, the SEED, SEED\_IV, and DEAP datasets served as the primary data sources for training models and assessing cross-subject emotion detection performance. To thoroughly assess the effectiveness of the proposed model, two distinct cross-validation strategies were adopted. These methodologies aimed to provide a comprehensive evaluation of the model's performance across various subjects.

We adopted a commonly used strategy for cross-subject affective detection. In this setup, the EEG data of one subject is designated as the $D_t$, while the combined data from all other subjects form the $D_s$. This procedure is repeated iteratively, allowing each subject to be tested as the $D_t$.

\begin{table}
	\centering
	\caption{The performance of representation methods on DEAP datasets using Cross-Subject Leave-One-Subject-Out Cross-Validation}
	\setlength{\tabcolsep}{1mm}
	\label{tab4}
	\begin{tabular}{c|c|c}
		\hline
		Method & Valance Pcc(\%) & Arousal Pcc(\%) \\
		\hline
		TSVM*~\cite{li2018bi}     & 61.77$\pm$08.93 & 56.59$\pm$11.98 \\\hline
		TPT*~\cite{li2018cross}   & 57.43$\pm$14.54 & 54.76$\pm$12.48 \\\hline
		TCA*~\cite{li2020novel}   & 56.23$\pm$14.33 & 51.81$\pm$15.03 \\\hline
		KPCA*~\cite{li2018cross}  & 54.35$\pm$10.22 & 58.15$\pm$14.96 \\\hline
		\textbf{SDA-DDA}          & 61.44$\pm$07.15 & 62.86$\pm$10.58 \\\hline
	\end{tabular}
\end{table}

Additionally, to thoroughly assess the model’s performance across sessions for each subject, we adopted a cross-subject, cross-session leave-one-subject-out cross-validation approach. In each iteration, all session data from one subject served as the $D_t$, while data from the remaining subjects’ sessions formed the $D_s$. This iterative approach ensured that every subject contributed as the $D_t$ at least once, enabling the calculation of average detection performance across all subjects.

\subsection{Results}
In Tables~\ref{tab2}, \ref{tab3} and \ref{tab4}, we conducted a comprehensive evaluation of various representations on the SEED, SEED\_IV and DEAP datasets using the leave-one-subject-out cross-validation method with a cross-subject single-session protocol. Our method demonstrated significant performance advantages on the SEED dataset, achieving an accuracy of 87.27\%$\pm$ 07.55\%. Similarly, on the SEED\_IV dataset, our method exhibited competitive accuracy of 74.01\%$\pm$11.34\%. In the DEAP dataset, SDA\_DDA achieves 61.44$\pm$07.15 for valence and 62.86$\pm$10.58 for arousal. These results strongly validate the substantial performance improvements achieved by our method on both datasets, surpassing the field average and demonstrating notable potential in the field of emotion recognition tasks. These findings provide compelling evidence supporting the effective application of our method in real-world scenarios.
\begin{table}[t]
	\centering
	\caption{The performance of shared methods on SEED and SEED\_IV datasets using Cross-Subject Cross-Session Leave-One-Subject-Out Cross-Validation}
	\setlength{\tabcolsep}{1mm}
	\label{tab_combined}
	\begin{tabular}{c|c|c}
		\hline
		Method & SEED  & SEED\_IV \\
		\hline
		KNN*~\cite{coomans1982alternative} & 60.66$\pm$07.93 & 40.83$\pm$07.28 \\\hline
		SA*~\cite{fernando2013unsupervised} & 61.41$\pm$09.75 & 64.44$\pm$09.46 \\\hline
		GFK*~\cite{li2018cross} & 66.02$\pm$07.59 & 45.89$\pm$08.27 \\\hline
		TCA*~\cite{pan2010domain} & 64.02$\pm$07.96 & 56.56$\pm$13.77 \\\hline
		CORAL*~\cite{sun2016return} & 68.15$\pm$07.83 & 49.44$\pm$09.09 \\\hline
		RF*~\cite{breiman2001random} & 69.60$\pm$07.64 & 50.98$\pm$09.20 \\\hline
		SVM*~\cite{suykens1999least} & 68.15$\pm$07.38 & 51.78$\pm$12.85 \\\hline
		DANN*~\cite{JMLR:v17:15-239} & 81.08$\pm$05.88 & 54.63$\pm$08.03 \\\hline
		\textbf{SDA-DDA} & \textbf{80.78\%$\pm$06.12\%} & \textbf{69.55\%$\pm$08.14} \\\hline
	\end{tabular}
\end{table}

Another crucial consideration for emotion brain-computer interfaces is the substantial variability observed among different subjects across various sessions. The evaluation approach of cross-subject and cross-session  poses a significant challenge for EEG-based emotion recognition models, requiring robust techniques for effective adaptation. To further validate this detection approach, which connect more closely with real-world application scenarios, we conducted experiments and obtained outstanding three-class classification performance on the SEED dataset, achieving an accuracy of 80.78\%$\pm$06.12\% (see Table~\ref{tab_combined}). Additionally, on the SEED-IV dataset, our model achieved a four-class accuracy of 69.55\%$\pm$08.14\% (see Table~\ref{tab_combined}). Compared to existing research, the proposed SDA-DDA method demonstrated industry-leading performance with a smaller standard deviation. These results indicate that the proposed SDA-DDA method exhibits excellent stability and generalization capabilities in handling subject and session differences.

\subsection{Confusion matrices}
\begin{figure}[htbp]
	\centering
	\begin{minipage}{0.49\linewidth}
		\centering
		\includegraphics[width=0.9\linewidth]{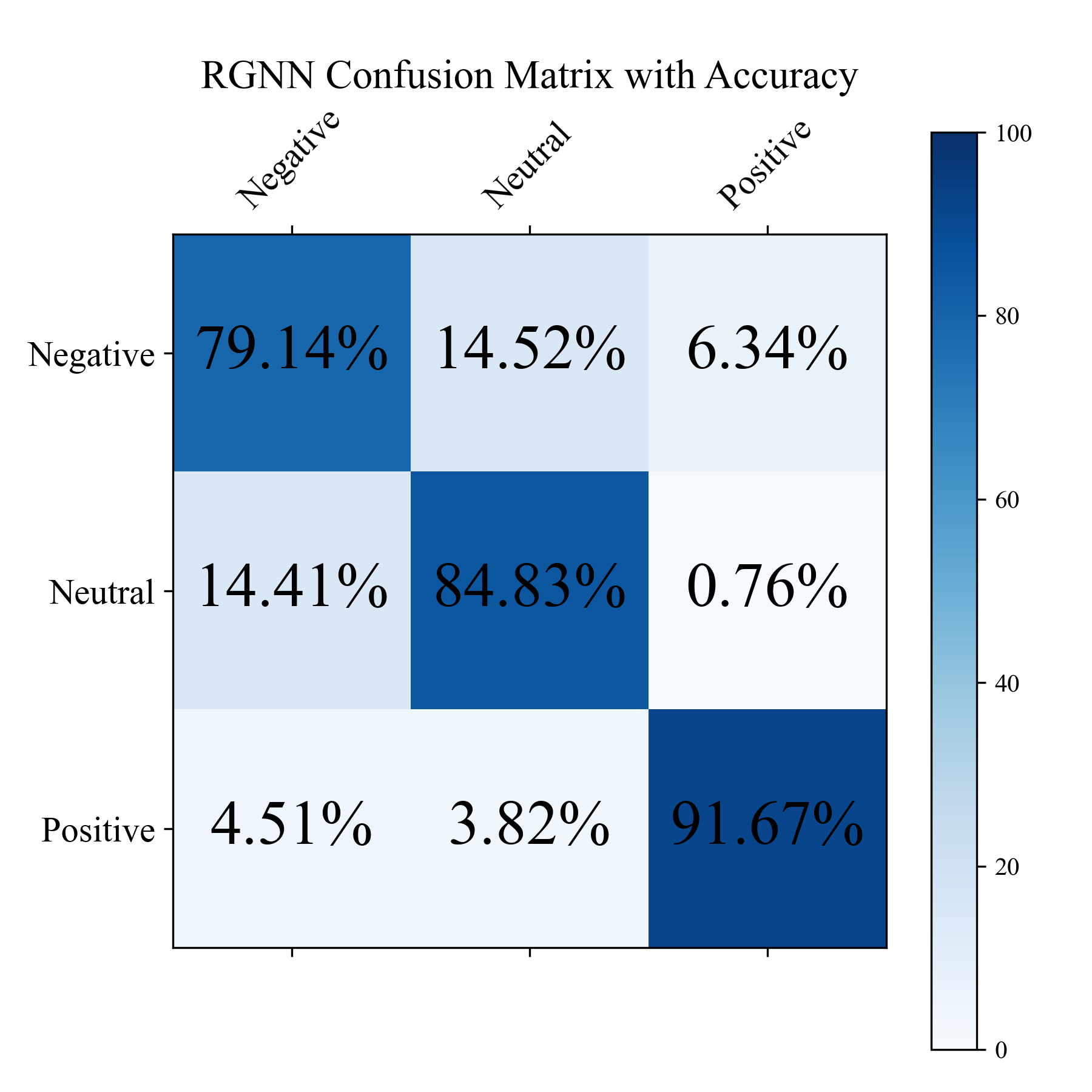}
		\label{fg:RGNN}
	\end{minipage}
	\begin{minipage}{0.49\linewidth}
		\centering
		\includegraphics[width=0.9\linewidth]{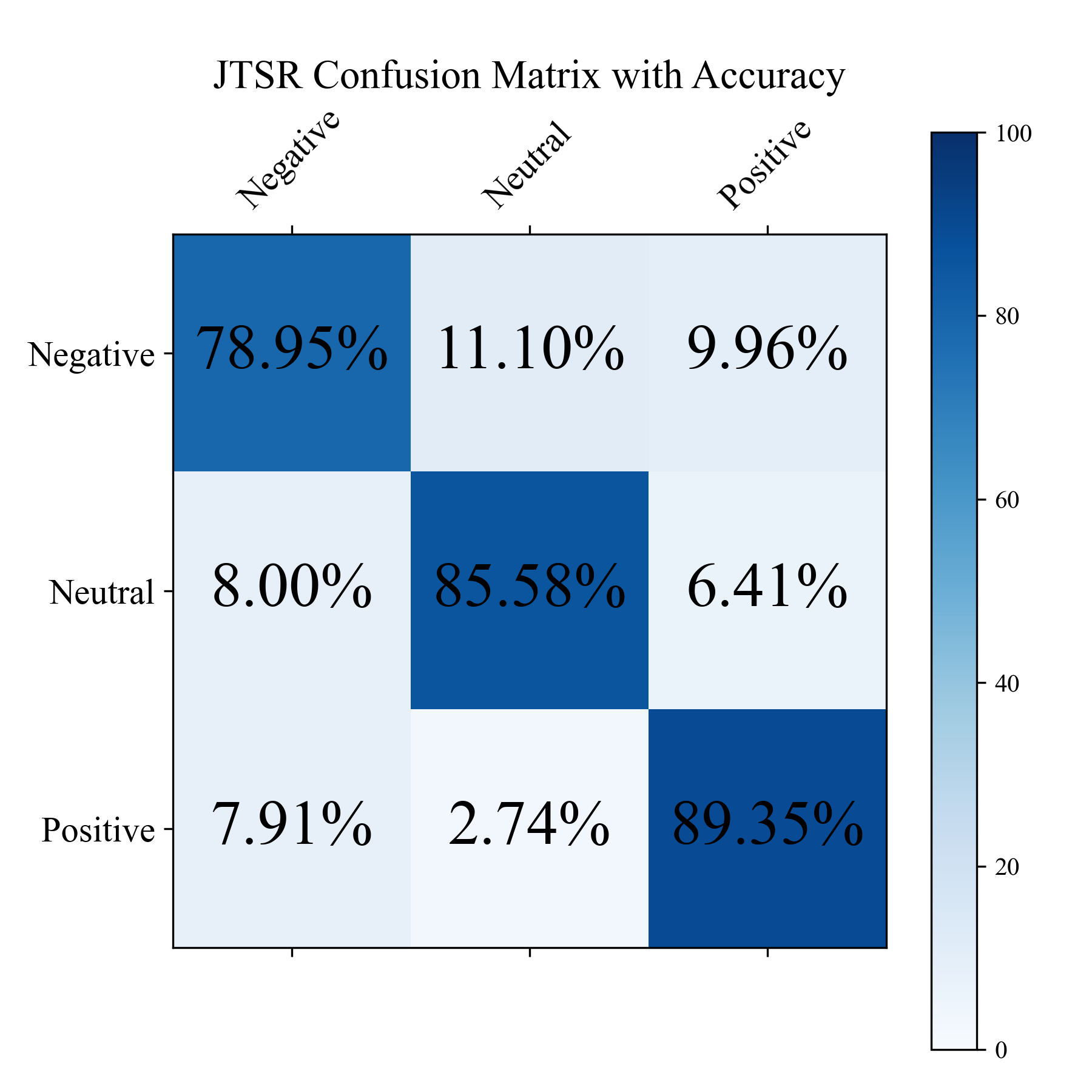}
		\label{fg:JTSR}
	\end{minipage}
	\begin{minipage}{0.49\linewidth}
		\centering
		\includegraphics[width=0.9\linewidth]{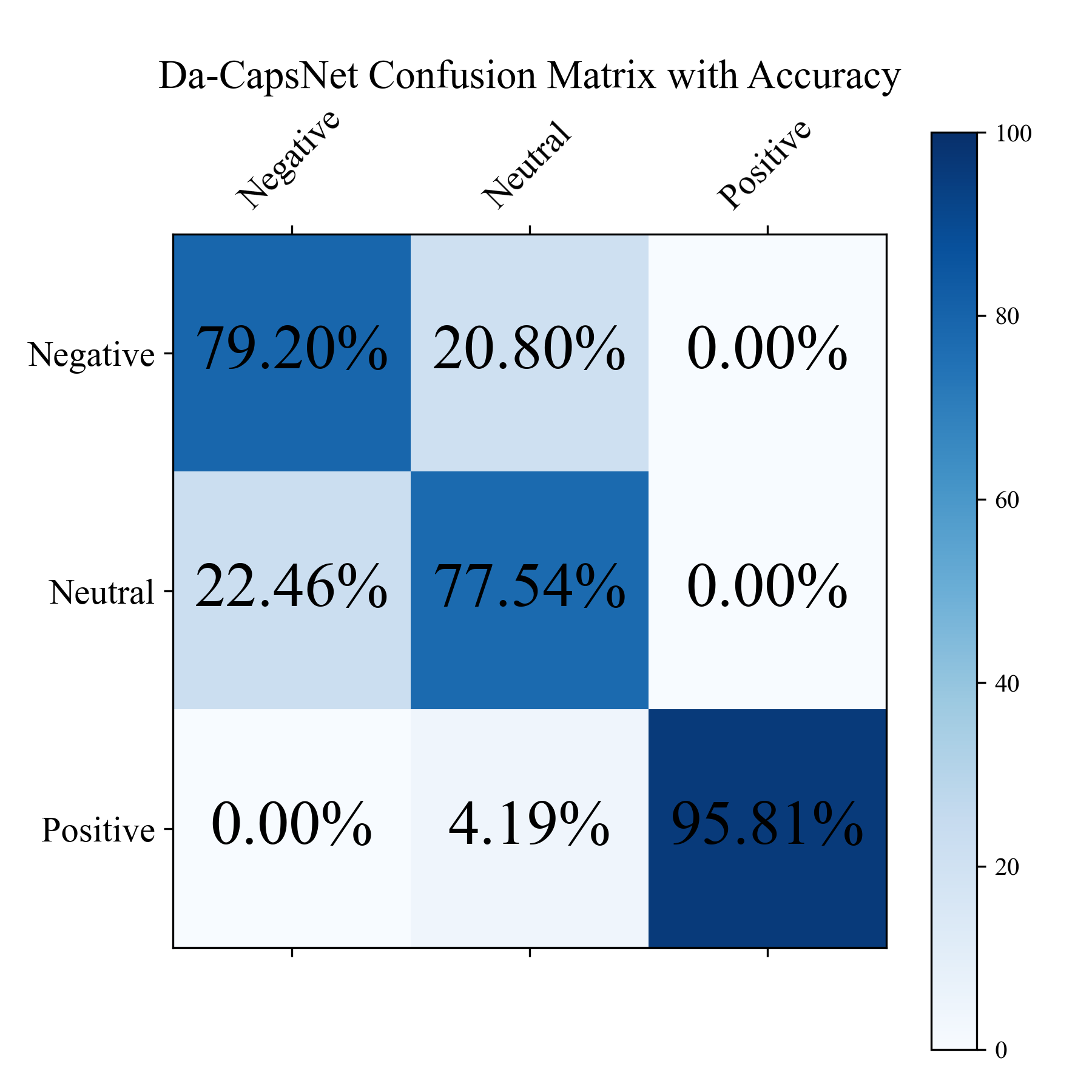}
		\label{fg:DA-CapsNet}
	\end{minipage}
	\begin{minipage}{0.49\linewidth}
		\centering
		\includegraphics[width=0.9\linewidth]{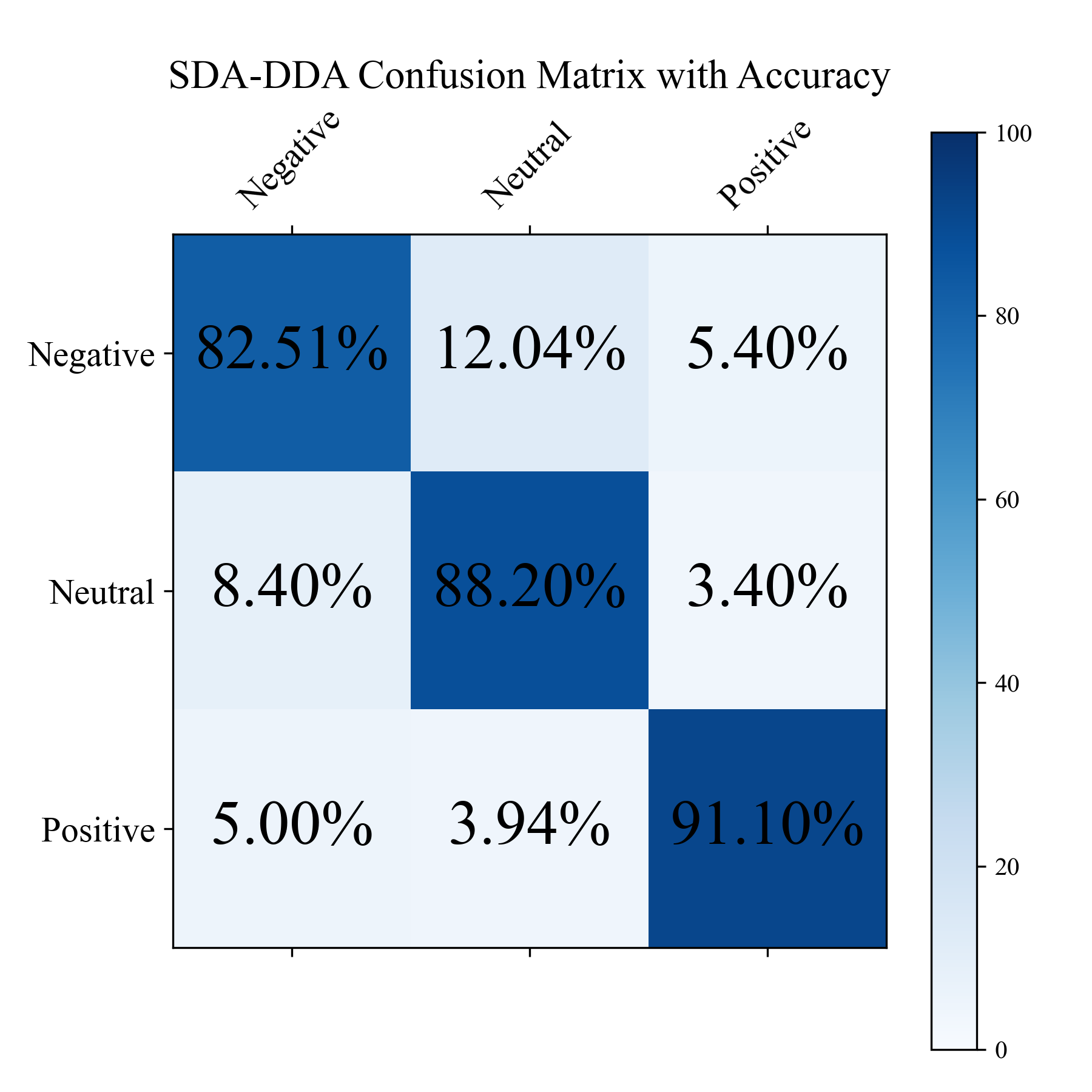}
		\label{fg:SDA-DDA}
	\end{minipage}
	\caption{Confusion matrices of different models:RGNN~\cite{zhong2020eeg}, JTSR~\cite{peng2022joint}, Da-CapsNet~\cite{liu2024capsnet} and SDA-DDA. The predicted labels are shown on the horizontal axis, and the true labels are represented on the vertical axis. \label{fg:ConfusionMatrix}}
\end{figure}

To comprehensively evaluate the performance of the proposed model across various emotional categories, we constructed confusion matrices and conducted a thorough comparative analysis with relevant literature. As illustrated in Fig.~\ref{fg:ConfusionMatrix}, the model achieved impressive accuracy rates of 82.51\%, 88.20\%, and 91.10\% for classifying negative, positive, and neutral emotions, respectively. It is noteworthy that the model has superior performance in classifying positive emotions compared to negative and neutral ones. Despite a slightly lower accuracy in negative emotion classification (82.51\%) compared to the other two categories, it still surpasses the performance of existing affective detection algorithms. This highlights the model's effectiveness in recognizing negative emotions, showcasing its distinct advantage over alternative approaches in the field. These findings underscore the robustness of the model and its potential to outperform existing methods, particularly in the nuanced classification of emotions, thereby providing valuable insights for future advancements in emotion recognition research.

\subsection{Ablation study}
This section presents a comprehensive set of ablation experiments to assess the efficacy of the proposed model.

\subsubsection{Experimental Setup}
To ensure fairness and reproducibility, the model parameters used in the ablation experiments are kept consistent throughout. The ablation realization models are described as follows:

EXP1: The baseline model employs a feature extractor and classifier to process features from both the $D_s$ and $D_t$, focusing on classification in the $D_t$.

EXP2: This variant enhances the baseline model by integrating the MMD mechanism to reduce the MPD gap between the $D_s$ and $D_t$.

EXP3: Building upon EXP2, this model incorporates CMMD to minimize differences in the CPD between the $D_s$ and $D_t$.

EXP4: Building on the baseline model, this variant incorporates MPD alignment and CPD alignment but excludes the use of dynamic distribution weights.

EXP5: Extending the baseline model, this approach introduces $\alpha$MMD and $\beta$CMMD without involving confidence filtering.

EXP6: This model incorporates dynamic distribution alignment and a confidence filtering mechanism into the original model architecture.

These distinct experimental configurations enable a systematic exploration of the contributions of each component, Offering meaningful insights into the model's behavior across different scenarios.

\begin{table}
	\centering
	\caption{Performance of the SDA-DDA model in the ablation study}\label{tab6}
	\setlength{\tabcolsep}{5mm}
	\begin{tabular}{c|c}
		\hline
		Ablation Experiment Strategy & Pcc (\%) \\
		\hline
		EXP1 & 81.35 $\pm$ 03.94 \\
		EXP2 & 86.47 $\pm$ 08.32 \\
		EXP3 & 85.75 $\pm$ 06.59 \\
		EXP4 & 85.20 $\pm$ 05.80 \\
		EXP5 & 86.50 $\pm$ 07.90 \\
		EXP6 & 87.27 $\pm$ 07.55 \\
		\hline
	\end{tabular}
\end{table}

\subsubsection{Ablation Results}
The ablation experiments, detailed in Table 6, were performed to evaluate the impact of the individual components of the proposed model. The baseline model (EXP1) demonstrated an accuracy of 81.35\% $\pm$ 03.94\%. The introduction of MMD in EXP2 improved accuracy to 86.47\% $\pm$ 08.32\%, demonstrating the effectiveness of reducing the distribution gap. By incorporating CMMD in EXP3, the emphasis shifted towards reducing differences in CPD, leading to an accuracy of 85.75\% $\pm$ 06.59\%. The combination of MMD and CMMD in EXP4 (85.20\% $\pm$ 05.80\%) demonstrated a harmonious performance in aligning both global and conditional distributions. Substantial improvement can be observed in EXP5 (86.50\% $\pm$ 07.90\%), where weighted MMD and CMMD were introduced without involving confidence filtering. The comprehensive model, EXP6, which integrated MMD, CMMD, and confidence filtering, exhibited the highest accuracy at 87.27\% $\pm$ 07.55\%, demonstrating the effectiveness of the confidence filtering mechanism in optimizing model performance. In summary, these findings provide nuanced insights into the contributions of each component, illustrating their roles in enhancing model accuracy across diverse experimental configurations.
\begin{figure}[ht]
	\centering
	\includegraphics[width=0.9\linewidth]{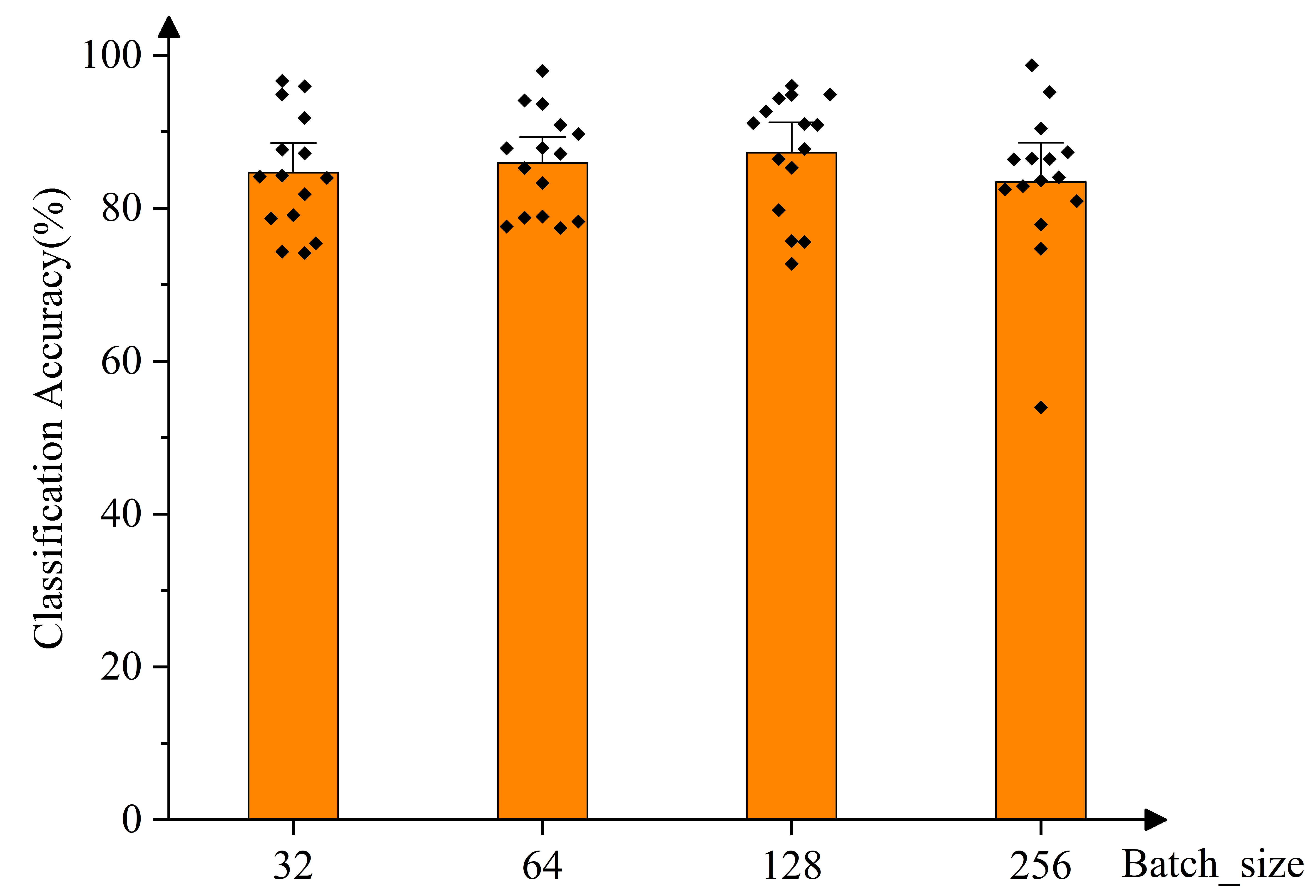}
	\caption{Average Accuracy by Batch Size for Each Experiment.}\label{fg:Figure_batch_size}
\end{figure}
\subsection{Sensitivity Analysis for Hyperparameters}
\subsubsection{Sensitivity analysis for batch-size and training epoch}
This section analyzes the model’s accuracy sensitivity to variations in batch size and training epochs, helping to identify configurations that maximize performance while maintaining stability.

As illustrated in Figure~\ref{fg:Figure_batch_size}, the model’s accuracy varies across different batch sizes (32, 64, 128, and 256). The analysis reveals that with a batch size of 32, the model achieves an average accuracy of approximately 84.66\% with a standard deviation of about 7.40\%, indicating relatively high variability and suggesting that smaller batches may lead to less stable training due to noisier gradient updates. Increasing the batch size to 64 improves the average accuracy slightly to around 85.91\%, while reducing the standard deviation to approximately 6.47\%, which points to improved stability as shown by narrower error bars. At a batch size of 128, the model reaches its highest average accuracy of approximately 87.27\%, with a standard deviation of about 7.56\%; this configuration appears optimal, achieving peak accuracy with moderate variability, as evidenced by relatively contained error bars. However, when the batch size is increased to 256, the average accuracy decreases to 83.43\% and the standard deviation rises to 9.82, resulting in increased variability and decreased stability, illustrated by wider error bars. This trend suggests that while larger batch sizes, such as 256, may offer computational efficiency, they can also lead to less precise gradient estimates, potentially degrading model performance. In summary, based on these observations from Figure~\ref{fg:Figure_batch_size}, a batch size of 128 is recommended as it maximizes accuracy while maintaining training stability, balancing performance and variability more effectively than both smaller and larger batch sizes.
\begin{figure}
	\centering
	\includegraphics[width=0.9\linewidth]{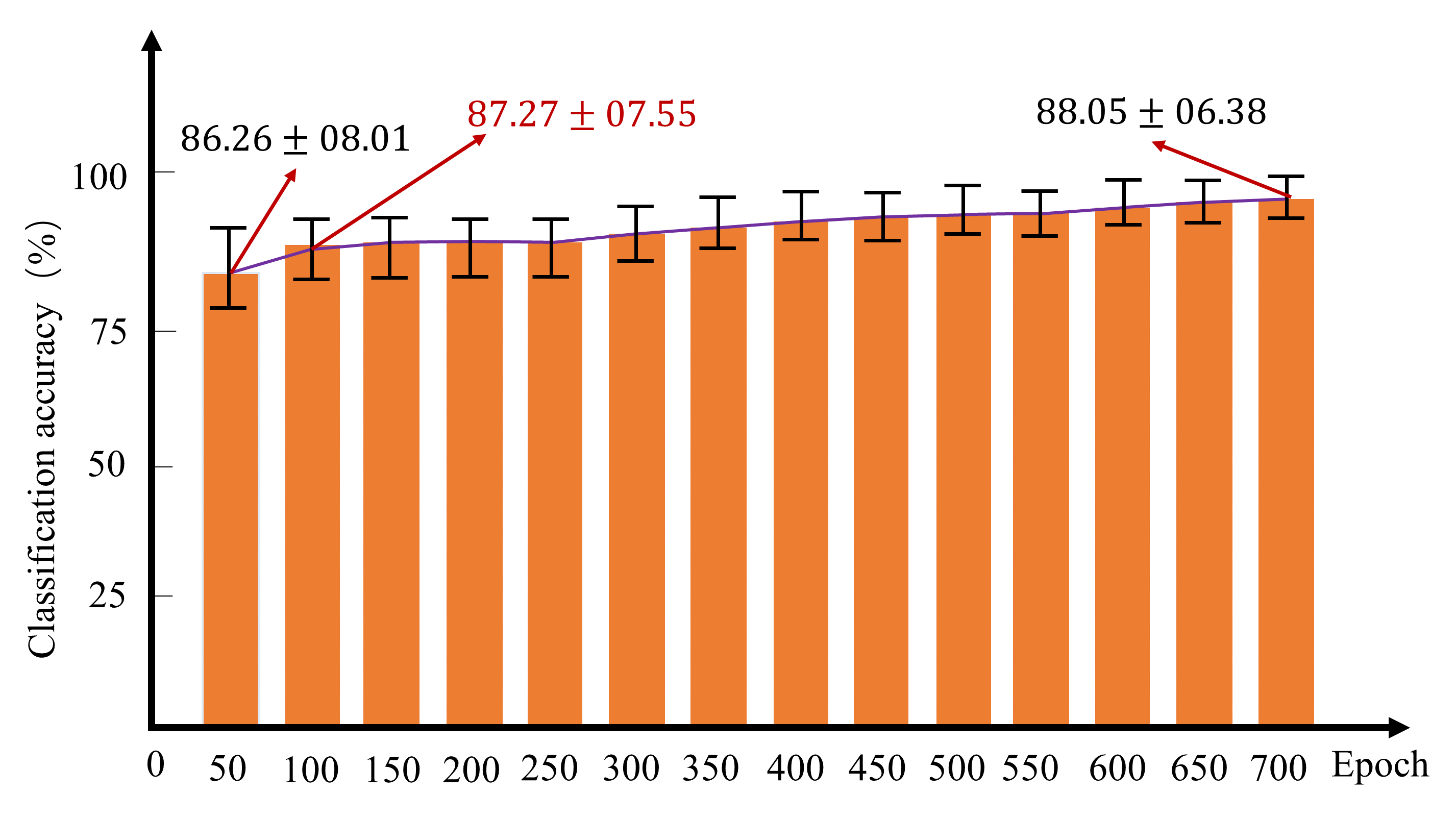}
	\caption{Effect of Epochs on Average Accuracy with Standard Deviatio.}\label{fg:Figure_epoch}
\end{figure}

As illustrated in Figure~\ref{fg:Figure_epoch}, we conducted a hyperparameter sensitivity analysis to assess the impact of training epochs on model performance. From epoch 50 to epoch 700, we recorded the model’s average accuracy and standard deviation (std) across 15 subjects to observe how performance evolves with extended training. The results indicate that increasing the number of epochs from 50 to 100 brings a notable improvement, with accuracy rising from approximately 86.26\% at epoch 50 to 87.27\% at epoch 100. Beyond this point, however, further increasing the epochs from 100 to 700—a sevenfold increase in training iterations—provides only a marginal accuracy gain of around 0.79, reaching a maximum of about 88.06\% at epoch 650. This finding suggests that the model reaches a high level of accuracy by epoch 100, with additional training iterations offering minimal performance enhancements. The consistently low standard deviation, which decreases slightly from 8.01 at epoch 50 to around 6.38 by epoch 650, indicates that the model's accuracy stabilizes with more epochs, but the improvements in accuracy are minimal. This observation highlights the model's relative insensitivity to further increases in the epoch parameter after 100 epochs, demonstrating rapid convergence to a high accuracy level within a relatively low number of iterations. Therefore, in practical applications, setting the epoch count to 100 can optimize computational resources by achieving efficient convergence without compromising accuracy, making it an optimal choice for balancing training efficiency and model performance.

\subsubsection{Sensitivity analysis for confidence}
\begin{figure}
	\centering
	\includegraphics[width=0.9\linewidth]{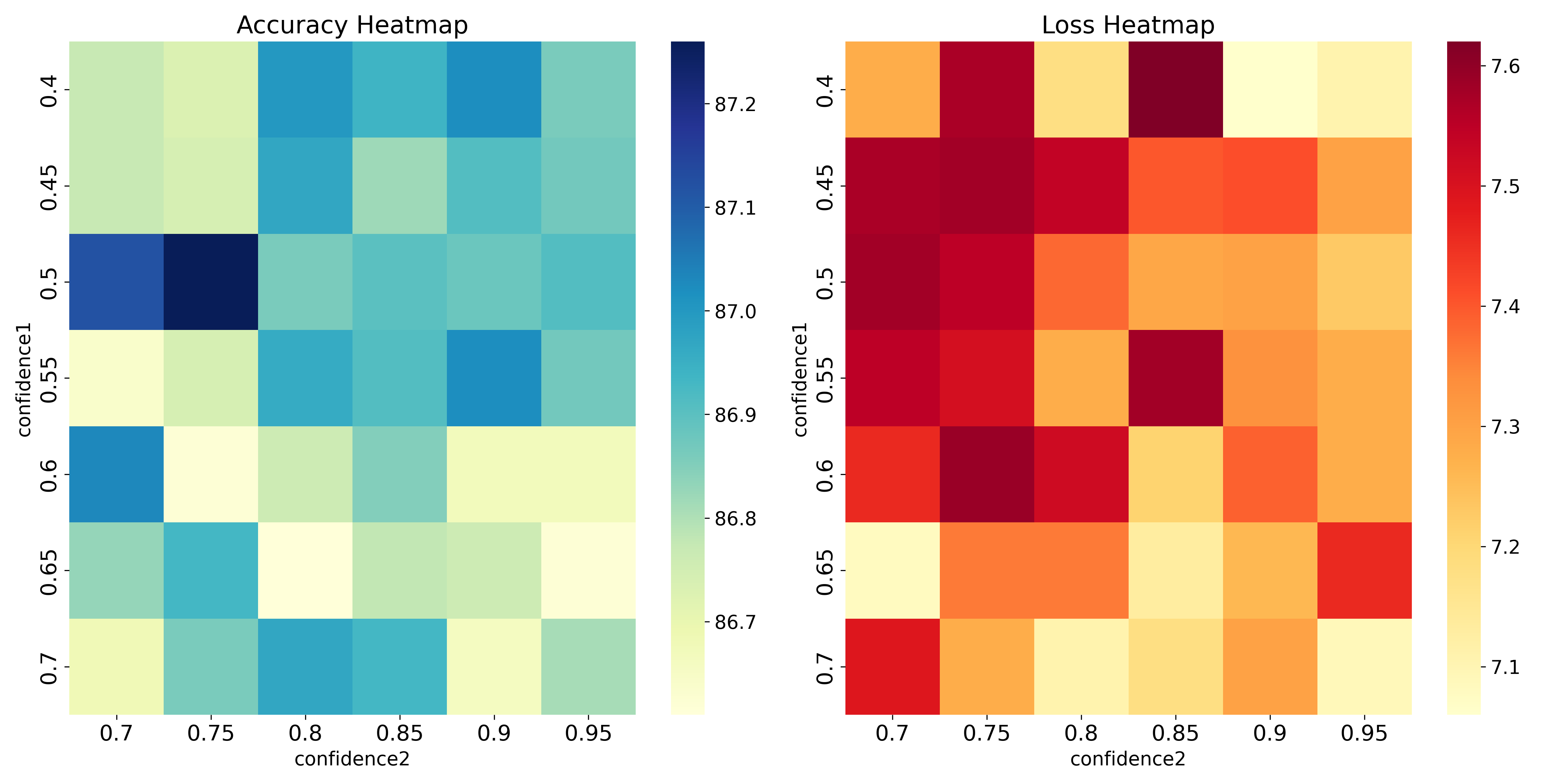}
	\caption{Effect of Epochs on Average Accuracy with Standard Deviation.}\label{fg:Figure_confidence}
\end{figure}

The confidence threshold settings in this study are illustrated in Figure 2. During the early phases of training, a higher proportion of $D_t$ samples is included to help the model gain a broader understanding of the $D_t$. During the training progress, it plays an important role to guiding the model to prioritize higher-confidence samples, thereby enhancing the stability and reliability of predictions. As training progresses, we gradually increase the confidence threshold. In the final stage, we set the confidence threshold to 1, meaning that only pseudo-labels with 100\% confidence are selected. This strict criterion is used to consolidate the model's performance by refining it with only the most reliable $D_t$ samples. At the last stage, the model is mature and has captured the essential patterns. Limiting training to the highest-confidence samples helps to prevent any potential degradation of performance.

We analyzed the model’s sensitivity to the confidence threshold hyperparameter by conducting a hyperparameter study over intermediate thresholds, with the batch size fixed at 128 and the number of training epochs set to 100. During this process, confidence1 and confidence2 correspond to different training phases, where confidence1 refers to the threshold between epochs 10 and 40, and confidence2 refers to the threshold between epochs 40 and 85. Specifically, confidence1 takes values between 0.4 and 0.7, while confidence2 takes values between 0.7 and 0.95. We performed a comprehensive grid search over these intermediate confidence thresholds and presented the average accuracy and standard deviation results across 15 subjects in Figure 7.
\begin{figure*}
	\centering
	\includegraphics[width=1\linewidth]{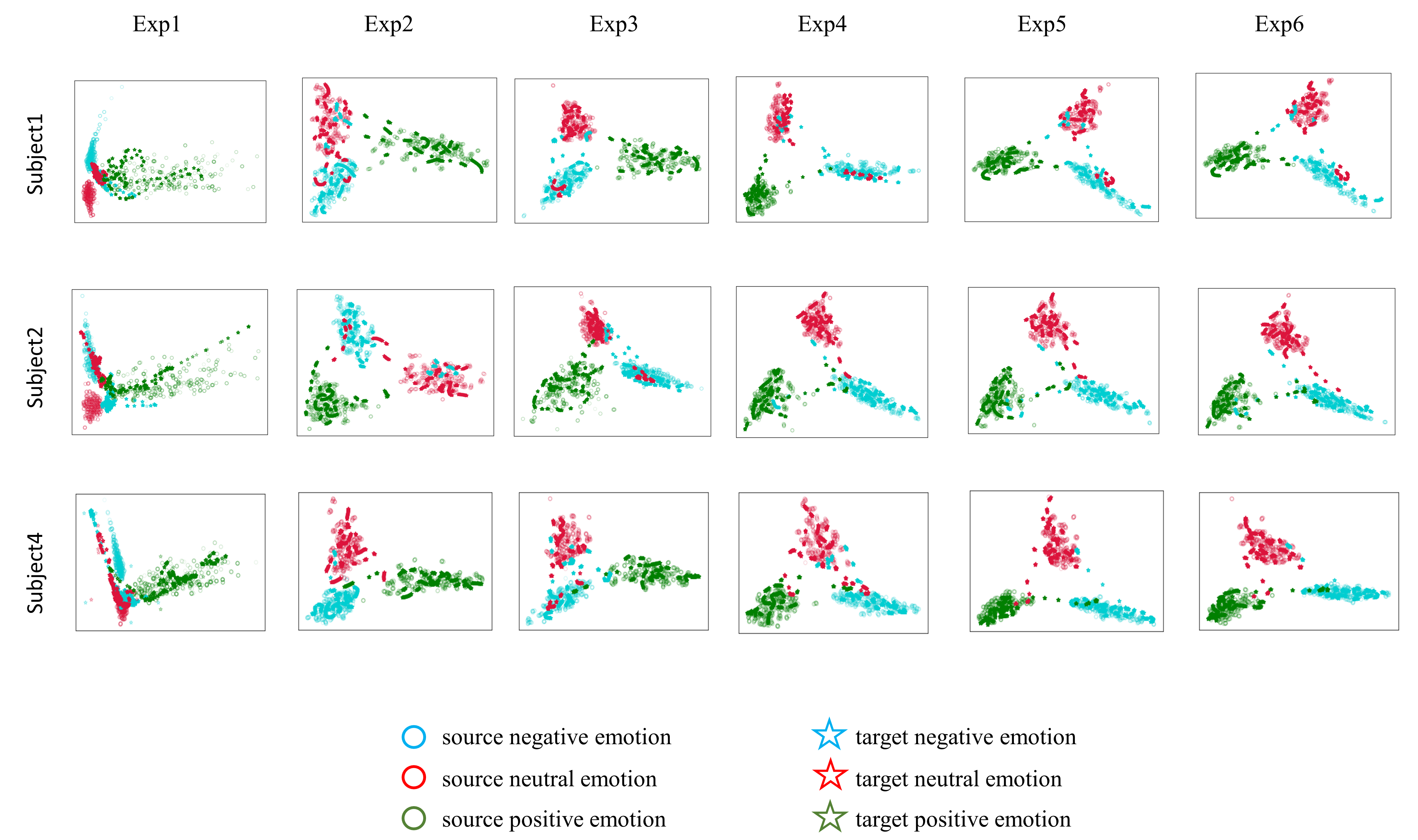}
	\caption{The t-SNE visualization illustrates the 2D embedding space of the learned features from both the source and target domains across six different model configurations.} \label{fg:t-SNE}
\end{figure*} 
Following the implementation of the confidence threshold mechanism, both the model’s accuracy and standard deviation improved significantly. For instance, by adjusting the combinations of confidence1 and confidence2, the highest accuracy reached 87.26\%, with the lowest standard deviation at 7.06. In contrast, under all combinations of confidence threshold settings, the model's accuracy and stability outperformed the scenario without any thresholds (Exp5: 86.5\% ± 7.9\%).This enhancement is primarily due to the confidence threshold mechanism's ability to effectively filter out low-quality pseudo-labels, thus avoiding interference from unreliable samples, leading to more stable model training and ultimately reaching better generalization performance in the $D_t$.

Through sensitivity analysis, we found that adjusting confidence1 and confidence2 within a reasonable range has a minor impact on model performance. Specifically, accuracy fluctuates between 86.62\% and 87.26\%, while the standard deviation ranges from 7.06 to 7.58.This result indicates that the selection of the confidence threshold is relatively insensitive to model performance; that is, the model can maintain good accuracy and stability across a broad parameter range. This robustness further confirms the effectiveness of the proposed confidence mechanism, allowing for greater flexibility in hyperparameter selection. It also means that in different application scenarios, strict tuning of confidence thresholds is not necessary to achieve stable performance.

\subsection{Representation Visualization}

Figure~\ref{fg:t-SNE} illustrates the experimental outcomes of six distinct model configurations, depicting the feature distributions learned by three participants in the source domain ($D_s$) and target domain ($D_t$) within a two-dimensional embedding space using t-SNE. These configurations correspond to the six strategies explored in the ablation study, each representing a unique method or adjustment for domain adaptation.

In EXP1, a preliminary clustering of data was observed, indicating a baseline level of similarity between $D_s$ and $D_t$. This suggests some degree of overlap between the domains, but with no active alignment mechanisms, the separation between the domains was still noticeable. When the MMD algorithm was introduced in EXP2, a more distinct clustering pattern emerged, demonstrating enhanced alignment of features between $D_s$ and $D_t$ by aligning the marginal probability distributions (MPDs). This improvement indicates that the MMD algorithm effectively narrows the gap between the domains by aligning their marginal distributions, resulting in a more coherent feature representation.

In EXP3, the integration of the CMMD algorithm alone caused the data from different categories to exhibit clearer separation, with distinct boundaries observed between categories. This suggests that aligning the conditional probability distributions (CPD) facilitates better differentiation of the emotional states across domains, reducing the impact of label noise in affective detection from EEG signals. The alignment of CPDs specifically aids in handling intra-class variations, allowing the model to more accurately segregate emotional states.

The combined application of both the MMD and CMMD algorithms in EXP4 resulted in a more refined feature distribution. Distinct classifications of different emotional categories were observed in both the source and target domains, further confirming the improved domain adaptation performance. By aligning both MPDs and CPDs, the model was able to optimize feature representations, enhancing its ability to transfer knowledge across domains.

In EXP5, the introduction of a weighted influence factor allowed for a more fine-grained optimization of the alignment process. By dynamically adjusting the influence of MPDs and CPDs during training, the model improved the overlap between source and target domains. The feature distributions in EXP5 showed a more unified representation across domains, suggesting that the dynamic weighting of domain-specific features significantly enhanced the alignment.

Finally, EXP6 introduced a pseudo-label confidence filtering mechanism, further refining the model's performance. This mechanism not only improved the confidence of pseudo-labels but also contributed to a more robust domain adaptation process by ensuring that the model focused on more reliable examples during training. The visualized data distributions in EXP6 demonstrate even greater coherence between the source and target domains, highlighting the positive impact of confidence filtering in improving feature overlap.

\subsection{Computational Efficiency and Inference Speed}
For real-time emotion recognition applications, a framework must maintain high efficiency and low latency to quickly respond to dynamic changes in emotional states. To assess the computational feasibility of our proposed method, we evaluated its floating-point operations  (FLOPs), parameter count, batch size, and inference speed on $D_t$. Our model demonstrated computational efficiency suitable for deployment on resource-limited devices, with a total computational load of 3.01 million FLOPs, a lightweight parameter count of only 0.01 million, and a batch size of 128 during testing. Notably, the model achieves an inference speed of approximately 20 milliseconds per batch on the $D_t$ data, underscoring its low latency. This combination of computational efficiency and rapid inference speed indicates that our model is promising candidate for real-time affective detection tasks in environments with limited computational resources.

\section{Discussion}\label{sec:discussion}

The variability between subjects in EEG data poses a significant challenge to the development of effective and widely applicable adaptive brain-computer interface models. To address this issue, we introduce a novel transfer learning framework called Semi-supervised Domain Adaptation with Dynamic Distribution Alignment. To assess the performance of SDA-DDA, we conducted experiments using the SEED, SEED\_IV, and DEAP datasets. Two validation strategies were employed to ensure a thorough evaluation of the framework’s effectiveness. We compared our method with several recently reported approaches, including machine learning-based methods (e.g., SVM, TKL, T-SVM, TCA, KPCA \cite{li2018bi, li2020novel, li2018cross}) and deep learning-based methods (e.g., RGNN, BiHDM, SA, MFA-LR, DA-CapsNet, MS-MDA, SCSTM-DS, MSFR-GCN~\cite{zhong2020eeg, li2020novel, jimenez2023learning, liu2024capsnet, chen2021ms, chen2023similarity, pan2023msfr}). The comparison results, summarized in Tables 2 through 6, highlight the superior performance of SDA-DDA. Our method combines conditional and marginal probability dynamic distribution alignment with a pseudo-label confidence threshold algorithm, achieving notable improvements in classification performance. To further validate the proposed method, we conducted additional studies, including ablation experiments, high-dimensional feature data visualization, and hyperparameter analysis. These analyses confirmed the strong performance of SDA-DDA, demonstrating its ability to effectively align distributions and enhance emotion classification accuracy.

Current domain adaptation methods for aligning emotional data distributions predominantly focus on marginal probability alignment{\cite{kouw2019review,guan2021domain,pan2023st,chen2021ms,he2022adversarial}}, with some models incorporating conditional probability alignment. However, these approaches often treat conditional and marginal probabilities as a combined joint probability distribution\cite{li2019domain,jiang2023generalization}. In this study, we propose a method leveraging MMD and CMMD modules to independently align MPD and CPD without increasing parameters or network complexity. Building on this, our dynamic distribution domain adaptation module comprehensively addresses the differences between these distributions during model training, achieving dynamic alignment of marginal and conditional probabilities. Traditional semi-supervised domain adaptation models for emotion recognition often neglect the impact of pseudo-label reliability on model performance throughout training\cite{Luo2021Progressive,Li2022Dynamic,Li2023Novel}. To address this, we introduce a pseudo-label confidence threshold module, significantly enhancing classification performance. Notably, our experiments confirm that this pseudo-label selection mechanism is robust to hyperparameter variations. To evaluate the impact of each component in the SDA-DDA framework, we conducted ablation studies. These experiments assessed the contributions of marginal probability alignment, conditional probability alignment, dynamic distribution alignment, and the pseudo-label confidence threshold module to the accuracy of emotion classification. Additionally, we performed high-dimensional data visualization using t-SNE, which provided compelling evidence that our method effectively separates source and target domains as well as their respective categories. Importantly, our model maintains a low computational cost and demonstrates minimal inference latency, making it a promising candidate for deployment on microcontrollers. This opens up opportunities for high-precision emotion recognition in real-world applications.

In the Future, several promising directions is deserved to research. Firstly, improving Pseudo-label Quality: One possible enhancement is adjusting the pseudo-label confidence threshold during the later stages of model training to ensure higher label quality. However, setting an excessively high threshold may discard too many uncertain pseudo-labels, leading to an imbalance between categories and low pseudo-label utilization\cite{zhang2021flexmatch, wang2022freematch}. Future work should focus on optimizing the balance between the quantity and quality of pseudo-labels, potentially through adaptive thresholds or other robust filtering techniques. Another important avenue is the real-time deployment of the SDA-DDA framework within an online Brain-Computer Interface system, facilitating dynamic user interaction. This would allow continuous model updates based on real-time EEG data, facilitating high-precision emotion recognition in dynamic, real-world settings. Testing and integrating our model into such an interactive online BCI environment would provide a strong foundation for personalized emotion recognition systems that can adapt in real-time.

\section{Conclusion}\label{sec:conclusion}
This paper introduces a novel framework for affective brain-computer interface classification, referred to as Semi-supervised Domain Adaptation with Dynamic Distribution Alignment. The framework addresses the challenges posed by data distribution differences, which hinder model accuracy, by aligning the marginal and conditional of the source and target domains. This is achieved through dynamic adjustment factors that balance the relative importance of these distributions. Additionally, a confidence filtering mechanism is incorporated to enhance the credibility of pseudo-labels, thereby improving classification performance. To validate the effectiveness of the proposed framework, extensive experiments were conducted using the SEED, SEED\_IV and DEAP datasets. Comparative analyses with existing methods demonstrate the superiority of SDA-DDA in terms of generalization and stability.

This contribution represents a significant advancement in affective brain-computer interfaces, with potential for further research and development. Future enhancements could enable this framework to address more complex transfer learning tasks in real-world scenarios.

\balance
\bibliographystyle{unsrt}
\bibliography{mybibfile.bib}

\end{document}